\documentclass{IEEEtran}
\usepackage{amsthm,amssymb,graphicx,amsmath,color,amsfonts,subcaption}\usepackage[update,prepend]{epstopdf}
\usepackage[noadjust]{cite}
\usepackage[latin1]{inputenc}
\usepackage{tikz}
\usetikzlibrary{arrows,calc}		
\usepackage{bbm} 
\usepackage{pdfpages}
\usepackage{tabulary}
\usepackage{multirow}
\usepackage{comment}

\def\nbo{{\mathbf{o}}}
\def\nbp{{\mathbf{p}}}

\def\nbu{{\mathbf{u}}}

\def\nbx{{\mathbf{x}}}
\def\nby{{\mathbf{y}}}

\def\nb0{{\mathbf{0}}}
\def\nb1{{\mathbf{1}}}

\def\nbX{{\mathbf{X}}}

\def\ncalA{{\mathcal{A}}}

\def\ncalC{{\mathcal{C}}}

\def\ncalV{{\mathcal{V}}}

\def\nbbE{{\mathbb{E}}}

\def\nbbP{{\mathbb{P}}}

\def\nbbR{{\mathbb{R}}}

\newtheorem{definition}{Definition}

\newtheorem{theorem}{Theorem}

\allowdisplaybreaks 

\begin{document}
 \title{Asymptotic Blind-spot Analysis of Localization Networks under Correlated Blocking using a Poisson Line Process}
\IEEEoverridecommandlockouts
\author{\IEEEauthorblockN{Sundar Aditya\IEEEauthorrefmark{1},~\IEEEmembership{Student~Member,~IEEE},
 Harpreet S. Dhillon\IEEEauthorrefmark{2},~\IEEEmembership{Member,~IEEE}, Andreas F. Molisch\IEEEauthorrefmark{1},~\IEEEmembership{Fellow,~IEEE},  and Hatim Behairy\IEEEauthorrefmark{3}}\\
\thanks{This work was supported by KACST under grant number 33-878.}
\thanks{S. Aditya and A. F. Molisch are with the Ming Hsieh Dept. of Electical Engineering, University of Southern California, USA. Email: \{sundarad, molisch\}@usc.edu}
\thanks{H. Dhillon is with the Bradley Dept. of Electical and Computer Engineering, Virginia Tech, USA. Email: hdhillon@vt.edu}
\thanks{H. Behairy is with the King Abdulaziz City for Science and Technology, Saudi Arabia. Email: hbehairy@kacst.edu.sa}
} 
\maketitle

\begin{abstract}
 In a localization network, the line-of-sight between anchors (transceivers) and targets may be blocked due to the presence of obstacles in the environment. Due to the {\em non-zero} size of the obstacles, the blocking is typically correlated across both anchor and target locations, with the extent of correlation increasing with obstacle size. If a target does not have line-of-sight to a minimum number of anchors, then its position cannot be estimated unambiguously and is, therefore, said to be in a {\em blind-spot}. However, the analysis of the blind-spot probability of a given target is challenging due to the inherent randomness in the obstacle locations and sizes. In this letter, we develop a new framework to analyze the worst-case impact of correlated blocking on the blind-spot probability of a typical target; in particular, we model the obstacles by a Poisson line process and the anchor locations by a Poisson point process. For this setup, we define the notion of the asymptotic blind-spot probability of the typical target and derive a closed-form expression for it as a function of the area distribution of a typical Poisson-Voronoi cell. As an upper bound for the more realistic case when obstacles have finite dimensions, the asymptotic blind-spot probability is useful as a design tool to ensure that the blind-spot probability of a typical target does not exceed a desired threshold, $\epsilon$.
\end{abstract} 
\begin{IEEEkeywords}
 Asymptotic blind-spot probability; Correlated blocking; Poisson Line process; Stochastic Geometry; Poisson-Voronoi tessellation
\end{IEEEkeywords}
\section{Introduction}
\begin{figure}
 \centering
 \includegraphics[scale=0.5]{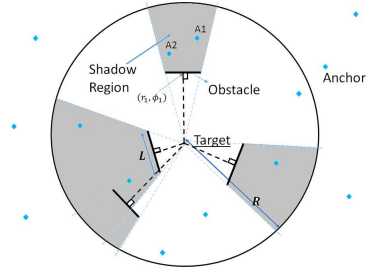}
 \caption{The unshadowed area (white region) surrounding a target is difficult to characterize due to the overlaps in the shadow regions caused by different obstacles.}
 \label{fig:small_L}
\end{figure}

Accurate localization is an important requirement for a variety of applications and in GPS-challenged environments (e.g. indoors), it is typically realized by deploying a network of transceivers, known as \emph{anchors}, over the region of interest (Fig. \ref{fig:small_L}). Depending on the localization technique used  (e.g., time-of-arrival (ToA), angle-of-arrival (AoA) etc.), a target should have line-of-sight (LoS) to at least a minimum number of anchors for unambiguous localization (e.g., for ToA-based localization over a 2D-plane, this number equals three). However, in many applications, the LoS link between a target and an anchor may be blocked by obstacles present in the environment. If a target does not have LoS to the required number of anchors, then a unique estimate of its position cannot be obtained and hence, is said to be in a \emph{blind-spot}.

If all the obstacle locations are known, then a deterministic, blind-spot eliminating placement of the anchors can be obtained by solving the art-gallery problem \cite{Gonzalez-Banos:2001:RAA:378583.378674}. On the other hand, if the obstacle locations are unknown, then the LoS blocking between a target and multiple anchors is a statistically dependent random phenomenon, where the extent of correlation is a function of the obstacle locations and sizes (e.g., in Fig. \ref{fig:small_L}, the blocking of anchors A1 and A2 to the target by a common obstacle induces correlation). A common assumption in the literature is to consider the blocking across different links to be mutually independent \cite{Schlo_Dhill_Buehr_2015, Bai_Vaze_Heath_2013}, which is reasonable for small obstacle sizes. However, for larger obstacles, this assumption can lead to the underestimation of a target's blind-spot probability. For instance, if two closely-spaced anchors are each blocked to a given target with probability $p$, then their joint blocking probability to the same target is also approximately $p$, which exceeds $p^2$, the result obtained by neglecting the blocking correlation and assuming independent blocking instead. Thus, it is important to consider the impact of obstacle-induced blocking correlation while analyzing a target's blind-spot probability.

In this paper, we characterize the worst-case impact of correlated blocking and introduce the notion of the \emph{asymptotic blind-spot probability} of the \emph{typical} target. In particular, we consider a stochastic geometry based approach by modeling the obstacles as a \emph{line process}, before deriving a closed-form expression for the asymptotic blind-spot probability of the typical target. Our approach is summarized below:
\begin{itemize}
 \item We assume the obstacles to be opaque to radio waves and model them using a Poisson line process (PLP) in $\nbbR^2$, where the projection of the origin, $\nbo$, onto the lines forms a homogeneous Poisson point process (PPP). This, in turn, induces a random polygon tessellation of $\nbbR^2$.
 \item We then show that the area distribution of the  polygon enclosing the typical target coincides with that of a typical Poisson-Voronoi cell.
 \item For anchors deployed according to a homogeneous PPP, we derive a closed-form expression for the asymptotic blind-spot probability of the typical target, using a well-known approximation for the area distribution of a typical Poisson-Voronoi cell.
 \end{itemize}
To the best of our knowledge, this is the first work that studies the impact of worst-case correlated blocking. The resulting expression that we obtain for a typical target's asymptotic blind-spot probability is an upper bound for the more realistic scenario when the obstacles have finite dimensions. Thus, our analysis provides useful design insights, such as the intensity with which anchors need to be deployed so that the blind-spot probability of a typical target does not exceed a desired threshold, $\epsilon$.

\section{System Model}  
\label{sec:sysmodel}
For our analysis, we focus on ToA-based localization over $\nbbR^2$. The extension to $\nbbR^3$ as well as other localization methods follows in a similar manner. Consider a network of anchors, each having a communication range, $R$, deployed over $\nbbR^2$ according to a homogeneous PPP, $\nbX_{\rm a}$, of intensity $\lambda$. 
\begin{figure}
 \centering
  \includegraphics[scale=0.53]{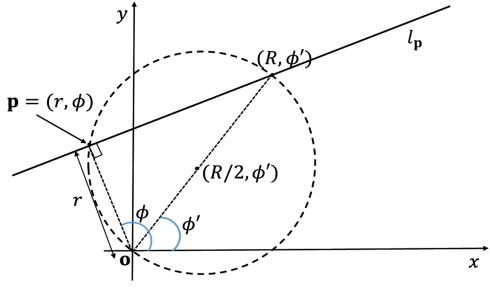}
  \caption{If the projection, $\nbp$, of $\nbo$ onto a line $l_{\nbp}$ lies inside the above disk, then $l_{\nbp}$ intersects the diameter from $\nbo$ to the point $(R,\phi')$.}
  \label{fig:3ints}
\end{figure}

We assume the obstacles to be opaque to radio signals and thus, it is convenient to imagine them as line-segments, since the obstacle thickness does not influence the LoS blocking between two points. For this obstacle shape, it is intuitive that the blocking correlation increases with line-segment length. Thus, to characterize the worst-case impact of correlated blocking on a target's blind-spot probability, we use lines to model the obstacle shapes in this work. As discussed in Section \ref{sec:asymp}, this enables us to obtain a useful performance bound for the more realistic, but less tractable, scenario of finite-sized obstacles. In Cartesian coordinates, a line in $\nbbR^2$ can be expressed as follows:
\begin{align}
 \label{eq:line}
 x\cos \phi + y \sin \phi = r
\end{align}
where $\nbp=(r,\phi)$ denotes the projection of $\nbo$ onto the line (e.g., Fig. \ref{fig:3ints}) in polar coordinates (i.e., $r\in[0,\infty), \phi \in [0,2\pi)$). It is easily seen that $\nbp$ uniquely determines a line in $\nbbR^2$. 

To capture the randomness in the obstacle locations, we model the obstacles using a line process, denoted by $\nbX_{l_{\nbp}}$. From (\ref{eq:line}), it can be seen that this gives rise to a point process, $\nbX_{\nbp}$, formed by the set of points, $\{\nbp\}$, which are the projections of $\nbo$ onto the lines in $\nbX_{l_{\nbp}}$. For the sake of tractability, we assume that $\nbX_{\nbp}$ forms a homogeneous PPP of intensity $\lambda_0$ over $\nbbR^2$, which results in $\nbX_{l_{\nbp}}$ forming a PLP, where each pair of lines has a unique point of intersection, with probability one. Consequently, $\nbX_{l_{\nbp}}$ splits $\nbbR^2$ into a collection of non-overlapping convex polygons, denoted by $\ncalC(\nbX_{l_{\nbp}})$, that form a tessellation of $\nbbR^2$. Let $C_\nbo(\lambda_0) \in \ncalC(\nbX_{l_{\nbp}})$ denote the polygon containing $\nbo$ and let $D_{\nbo}(R)=\{(r,\phi): r\in [0,R], \phi \in [0,2\pi)\}$ denote the disk of radius $R$, centered at $\nbo$. 

\begin{figure} 
\centering
 \includegraphics[scale=0.3]{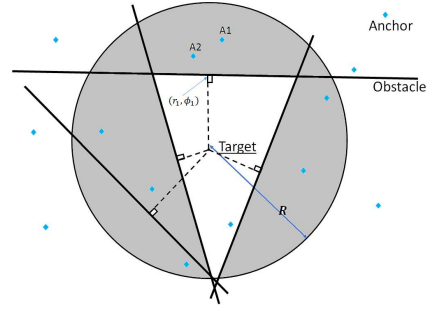}
 \caption{If the obstacles are lines, then the unshadowed region is a convex region. In particular, it is a convex polygon if no point at a distance $R$ from the target has LoS to it.}
 \label{fig:large_L}
\end{figure}

Due to the stationarity of the anchor PPP, it can be assumed without loss of generality that a target is situated at $\nbo$, which we refer to as the \emph{typical target}. Based on our model, the anchors having LoS to the typical target are constrained to lie in $C_\nbo(\lambda_0) \cap D_{\nbo}(R)$ (Fig. \ref{fig:large_L}). Hence, the typical target is in a blind-spot if and only if there are fewer than three anchors present in $C_\nbo(\lambda_0) \cap D_{\nbo}(R)$. The special case when $C_\nbo(\lambda_0) \subseteq D_{\nbo}(R)$ is of particular interest as the blind-spot probability of the typical target depends only on the area distribution of $C_\nbo(\lambda_0)$ (as shall be seen in Section \ref{sec:asymp}), for which accurate closed-form approximations exist. It is intuitive that as $\lambda_0$ increases, $C_\nbo(\lambda_0) \subseteq D_{\nbo}(R)$ with high probability (Fig. \ref{fig:large_L}). To formalize this notion, let
\begin{align}
 v(r,\phi)&= \begin{cases}
 			  & 1, ~ \mbox{ if the point $(r,\phi)$ has LoS to $\nbo$} \\
 			  & 0, ~ \mbox{ else}.
 			 \end{cases}
\end{align}
If $C_\nbo(\lambda_0) \nsubseteq D_{\nbo}(R)$, then there exists at least one direction $\phi'\in[0,2\pi)$ such that $v(R,\phi')=1$. For this condition to be satisfied, no point from $\nbX_{\nbp}$ should lie in a disk of diameter $R$, centered at $(R/2,\phi')$ (see Fig. \ref{fig:3ints}). Therefore, $\nbbP(v(R,\phi')=1)=\exp(-\lambda_0 \pi R^2/4)$, for any $\phi' \in [0,2\pi)$. Hence, for an arbitrarily small $\delta \in (0,1)$ such that $\exp(-\lambda_0 \pi R^2/4) < \delta$, $C_\nbo(\lambda_0) \subseteq D_{\nbo}(R)$ with probability greater than $1-\delta$. Therefore, we develop our analysis under the assumption that $C_\nbo(\lambda_0) \subseteq D_{\nbo}(R)$ by considering a sufficiently large value of $\lambda_0$, determined by the parameter $\delta$. 

\section{Asymptotic Blind-spot Probability}
\label{sec:asymp}
 Let $A_v$ denote the area of $C_\nbo(\lambda_0)$ and $g(A_v;\lambda,\lambda_0)$ the blind-spot probability of the typical target, conditioned on $A_v$, with parameters $\lambda$ and $\lambda_0$. Then,
\begin{align}
 \label{eq:g_wc}
 g(A_v;\lambda,\lambda_0)&= \displaystyle\sum\limits_{k=0}^2 \nbbP(k \mbox{ anchors present in } C_\nbo(\lambda_0)) \notag \\
 &= e^{-\lambda A_v}\left(1+\lambda A_v + \frac{(\lambda A_v)^2}{2}\right).
\end{align}
\begin{definition}
The asymptotic blind-spot probability of the typical target, denoted by $b_{\rm as}(\lambda,\lambda_0)$, is defined as follows: 
\begin{align}
 \label{eq:pbs_wc}
 b_{\rm as}(\lambda,\lambda_0)&\triangleq \displaystyle\int\limits_0^\infty g(A_v;\lambda,\lambda_0) f(A_v) {\rm d}A_v
\end{align}
where $f(A_v)$ denotes the probability density function (pdf) of $A_v$.
\end{definition}
Thus, $b_{\rm as}(\lambda,\lambda_0)$ is the unconditional blind-spot probability of the typical target, due to $\nbX_{l_{\nbp}}$ and $\nbX_{\rm a}$, and depends on the area distribution of $C_\nbo(\lambda_0)$. In particular, $f(A_v)$ fully captures the worst-case impact of correlated blocking as the obstacles are \emph{infinitely long}. We now proceed to characterize $f(A_v)$, for which we define the following terms, before stating our main result in Theorem \ref{thm:main}. Although the result presented in Theorem \ref{thm:main} exists in the literature on random polygon tessellations (see \cite{Hilhorst_Calka_2008}), it has, to the best of our knowledge, never been applied to a localization setting previously. Since its proof is quite straightforward, we include it for completeness. 
\begin{definition}[Voronoi cell]
\label{def:vor}
For a countable set of points $\ncalA \subseteq\nbbR^2$, the Voronoi cell of $\nbx\in \ncalA$, denoted by $\ncalV_{\nbx}(\ncalA)$, is defined as follows:
\begin{align}
 \ncalV_{\nbx}(\ncalA)=\{\nby\in \nbbR^2: \|\nby-\nbx\|_2 \leq \inf_{\nbu \in \ncalA \setminus \nbx } \|\nby-\nbu\|_2\}
\end{align}
where $\|.\|_2$ denotes the $L_2$-norm. In other words, $\ncalV_{\nbx}(\ncalA)$ contains all the points in $\nbbR^2$ that are closer to $\nbx$ than any other point in $\ncalA$.

\end{definition}

\begin{definition}[Poisson-Voronoi cell]
\label{def:Pois_vor}
If $\ncalA$ is a realization of a homogeneous PPP of intensity $\mu$, then $\ncalV_\nbx(\ncalA)$ is referred to as a Poisson-Voronoi cell with parameter $\mu$, for $\nbx \in \ncalA$. In particular, the expected area of $\ncalV_\nbx(\ncalA)$ equals $1/\mu$.
\end{definition}
 
\begin{figure}
  \centering
 \includegraphics[scale=0.25]{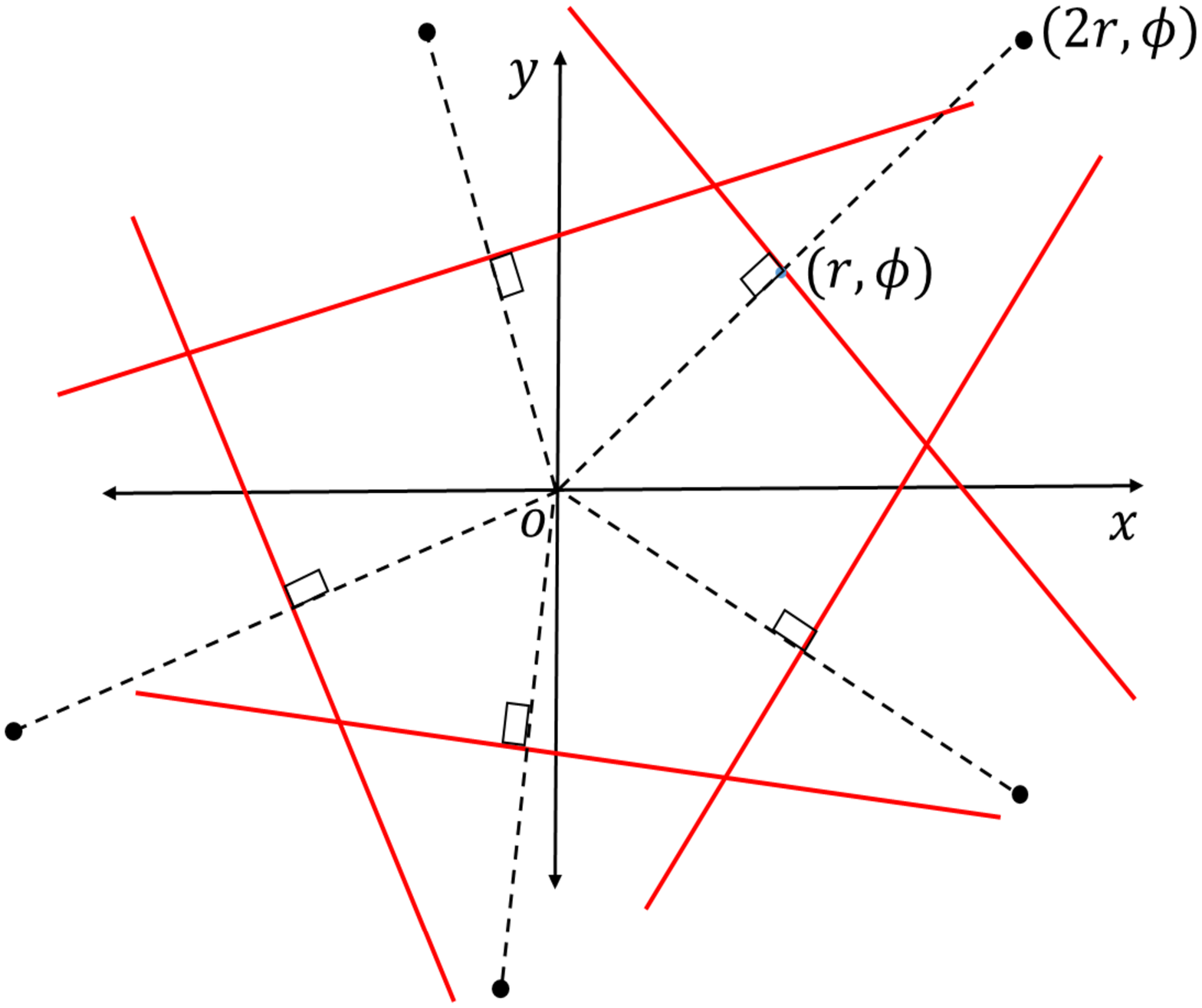}
 \caption{The interior of the polygon, $C_{\nbo}(\lambda_0)$, surrounded by the red lines is the Voronoi cell, $\ncalV_\nbo(\ncalA)$, where $\ncalA$ contains $\nbo$ and the points shown in black circles.}
 \label{fig:proofdiag}
\end{figure}

\begin{theorem}\label{thm:main}
 The area distribution of $C_\nbo(\lambda_0)$ coincides with that of a typical Poisson-Voronoi cell with parameter $\lambda_0/4$.
\end{theorem}
\begin{IEEEproof}
The point process $2\nbX_{\nbp}$ is a homogeneous PPP of intensity $\lambda_0/4$. By the Slivnyak-Mecke Theorem \cite{Sto_et_al_full_2013}, $\ncalA=2\nbX_{\nbp} \cup \{\nbo\}$ has the same distribution as $2\nbX_{\nbp}$. Hence, from Definition \ref{def:Pois_vor}, $\ncalV_{\nbo}(\ncalA)$ is a Poisson-Voronoi cell with parameter $\lambda_0/4$. By construction, $\ncalV_{\nbo}(\ncalA)$ coincides with $C_{\nbo}(\lambda_0)$, as illustrated in Fig. \ref{fig:proofdiag}. Therefore, $C_{\nbo}(\lambda_0)$ has the same area distribution as $\ncalV_{\nbo}(\ncalA)$.
\end{IEEEproof}
For a typical Poisson-Voronoi cell with parameter $\lambda_0/4$, the pdf of its area is well-approximated by a three-parameter Gamma distribution \cite{Hinde_Miles_1980, Tanemura_2003_statisticaldistributions}, given by:
\begin{align}
 \label{eq:Av_pdf}
  \hspace{-8mm} f(A_v)&= \begin{cases}
               & \hspace{-3mm}\frac{ab^{(c/a)}}{\Gamma(c/a)}\left(\frac{\lambda_0}{4}\right)^c A_v^{c-1}e^{-b(\lambda_0 A_v/4)^a}, A_v\geq 0 \\
               & 0, ~ \mbox{else}
               \end{cases}
\end{align}
where $a=1.07950$, $b=3.03226$, $c=3.31122$ and $\Gamma(z)=\int_0^\infty x^{z-1} e^{-x} {\rm d}x$ for $z>0$. Substituting (\ref{eq:Av_pdf}) in (\ref{eq:pbs_wc}) and evaluating the integral, we obtain $b_{\rm as}(\lambda,\lambda_0)$.

On the other hand, if we ignore correlated blocking and assume independent blocking, the unblocked anchors to the typical target form a point process obtained by independently sampling the anchor PPP, where the sampling probability of an anchor at $(r,\phi)\in D_{\nbo}(R)$ equals $\nbbP(v(r,\phi)=1)$. As a result, the unblocked anchors form a non-homogenous PPP \cite{Sto_et_al_full_2013} over $D_{\nbo}(R)$ whose intensity at the point $(r,\phi)$, denoted by $\lambda_{\rm ind}(r,\phi)$, is given by
\begin{align}
 \label{eq:indep_intensity}
 \lambda_{\rm ind}(r,\phi)= \lambda \mathbb{P}(v(r,\phi)=1) = \lambda e^{-\lambda_0 \pi r^2/4}.
\end{align}
For a non-homogeneous PPP with intensity $\lambda_{\rm ind}(r,\phi)$, the number of points in $D_{\nbo}(R)$ is a Poisson random variable with mean $\lambda \nbbE[A_v]$, given by
\begin{align}
 \lambda \nbbE[A_v] = \displaystyle\int\limits_0^{2\pi} \displaystyle\int\limits_0^R \lambda_{\rm ind}(r,\phi) r {\rm d}r {\rm d}\phi &= \frac{4\lambda}{\lambda_0}(1-e^{-\frac{\lambda_0 \pi  R^2}{4}}).
\end{align}
Therefore, the blind-spot probability of the typical target due to the independent blocking assumption, denoted by $b_{\rm as}^{\rm ind}(\lambda,\lambda_0)$, has the following expression, similar to (\ref{eq:g_wc}):
\begin{align}
\label{eq:pbs_indep}
 b_{\rm as}^{\rm ind}(\lambda,\lambda_0)&= e^{-\lambda \nbbE[A_v] }\left(1+ \lambda \nbbE[A_v] + \frac{(\nbbE[A_v])^2}{2}\right).
\end{align}

To model the realistic case of obstacles having finite dimensions, each line $l_{\nbp}\in \nbX_{l_{\nbp}}$ can be truncated to a line segment of length $L$ with mid-point $\nbp$, as shown in Fig. \ref{fig:small_L}. For this scenario, let $B_v$ denote the area of the unshadowed region in $D_{\nbo}(R)$, surrounding the typical target. Similar to (\ref{eq:g_wc}), the conditional blind-spot probability of the typical target, denoted by $g(B_v;\lambda,\lambda_0,L,R)$ with parameters $\lambda$, $\lambda_0$, $L$ and $R$, has the following expression, similar to (\ref{eq:g_wc}):
\begin{align}
 g(B_v;\lambda,\lambda_0,L,R)&= e^{-\lambda B_v}\left(1+\lambda B_v + \frac{(\lambda B_v)^2}{2}\right).
\end{align}
The unconditional blind-spot probability of the typical target, denoted by $b(\lambda,\lambda_0,L,R)$, is then given by:
\begin{align}
\label{eq:pbs_true}
 b(\lambda,\lambda_0,L,R)&=\displaystyle\int\limits_0^{\pi R^2} g(B_v;\lambda,\lambda_0,L,R) f(B_v) {\rm d}B_v
\end{align}
where $f(B_v)$ denotes the pdf of $B_v$ and in a manner similar to $f(A_v)$, captures the blocking correlation induced by obstacles of length $L$. However, unlike $f(A_v)$, a closed-form characterization of $f(B_v)$ is difficult to obtain due to the generally non-convex shape of the unshadowed region, which is further complicated by the overlaps in the shadow regions induced by the obstacles (Fig. \ref{fig:small_L}). However, since larger obstacles cause more blocking, the following inequality holds:
\begin{align}
 \label{ineq:len}
 b(\lambda,\lambda_0,l_1,R) &\leq b(\lambda,\lambda_0,l_2,R) \hspace{2mm} \mbox{for} \hspace{2mm} l_1 \leq l_2. 
\end{align}
Hence, for sufficiently large $\lambda_0$ such that $C_{\nbo}(\lambda_0) \subseteq D_{\nbo}(R)$ with high probability,
\begin{align}
  \label{eq:final}
 b(\lambda,\lambda_0,L,R) \leq \lim_{l \rightarrow \infty} b(\lambda,\lambda_0,l,R) = b_{\rm as}(\lambda,\lambda_0)
\end{align}
for any $L>0$. Therefore, $b_{\rm as}(\lambda,\lambda_0)$ can be used as a design tool to determine the anchor intensity required such that the blind-spot probability of a typical target is no more than a desired threshold, $\epsilon$. Finally, for completeness, if we assume independent blocking for the finite obstacle case, then the resulting blind-spot probability of the typical target, denoted by $b^{\rm ind}(\lambda,\lambda_0,L,R)$, has the following expression, similar to (\ref{eq:pbs_indep}):
\begin{align}
\hspace{-4mm} b^{\rm ind}(\lambda,\lambda_0,L,R) &= e^{-\lambda \nbbE[B_v] }\left(1+ \lambda \nbbE[B_v] + \frac{(\lambda \nbbE[B_v])^2}{2}\right) \notag \\
\label{eq:pbs_indep_asymp}
&\leq \lim_{l \rightarrow \infty} b^{\rm ind}(\lambda,\lambda_0,l,R)= b_{\rm as}^{\rm ind}(\lambda,\lambda_0)
\end{align} 
where the inequality in (\ref{eq:pbs_indep_asymp}) holds for any $L>0$.

\section{Numerical Results}
\begin{figure}
\centering
 \includegraphics[scale=0.43]{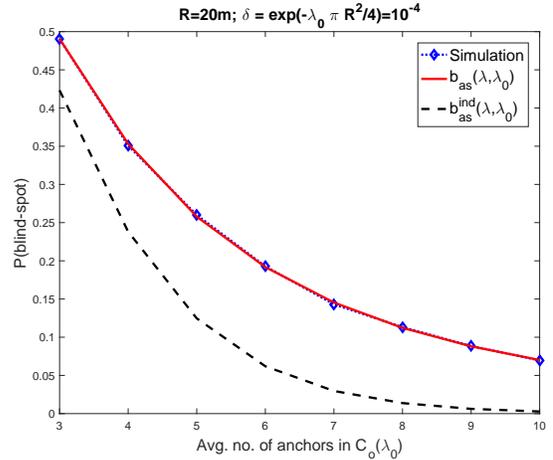}
 \caption{$b_{\rm as}(\lambda,\lambda_0)$ accurately chracterizes the blind-spot probability if there exists a sufficiently high intensity of `large' obstacles.}
 \label{fig:sim} 
\end{figure}

\begin{figure}
 \centering
 \includegraphics[scale=0.43]{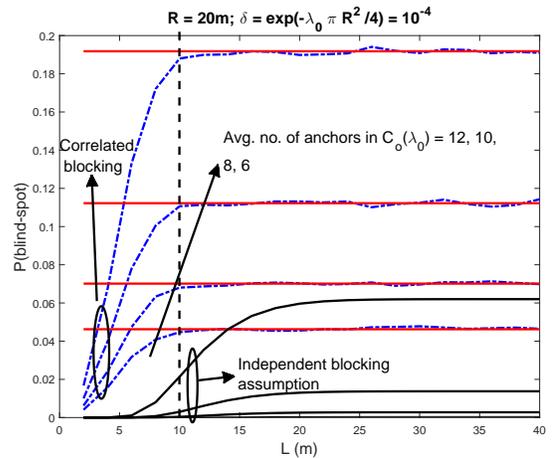}
 \caption{The dashed blue, solid red and solid black curves plot $b(\lambda,\lambda_0,L,R)$, $b_{\rm as}(\lambda,\lambda_0)$ and $b^{\rm ind}(\lambda,\lambda_0,L,R)$, respectively. For $L\geq R/2$, the line assumption for obstacles holds, given a sufficiently high intensity of obstacles.}
 \label{fig:Lcompare}
\end{figure}

We consider a circle of radius $R=20{\rm m}$ to be our region of interest. For $\delta=10^{-4}$, the plot of $b_{\rm as}(\lambda,\lambda_0)$ as a function of $\lambda$ is shown in Fig. \ref{fig:sim}, based on $10^{5}$ Monte Carlo simulations for each $\lambda$. The closed-form expression for $b_{\rm as}(\lambda,\lambda_0)$ obtained in (\ref{eq:pbs_wc}) mirrors the empirically observed results, thereby justifying the approximation that $C_{\nbo}(\lambda_0) \subseteq D_{\nbo}(R)$ for the chosen value of $\lambda_0$. On the other hand, $b_{\rm as}^{\rm ind}(\lambda,\lambda_0)$ underestimates $b_{\rm as}(\lambda,\lambda_0)$ since it ignores the blocking correlation. Furthermore, if $C_{\nbo}(\lambda_0) \subseteq D_{\nbo}(R)$, then clearly, $b(\lambda,\lambda_0,L,R)=b_{\rm as}(\lambda,\lambda_0)$ for $L\geq 2R$, since each obstacle forms a secant. However, we observe from Fig. \ref{fig:Lcompare} that the the bound in (\ref{eq:final}) is tight for much smaller obstacle lengths starting from approximately $R/2$. Additionally, we also observe in Fig. \ref{fig:Lcompare} that the independent blocking assumption reasonable for small $L$, but breaks down for larger $L$ as the extent of blocking correlation increases.

\section{Conclusion}
In this letter, we investigated the worst-case impact of correlated blocking on the blind-spot probability of a typical target in a localization network by assuming a PLP obstacle model, which induces a random polygon tessellation of the plane. We then defined the notion of the asymptotic blind-spot probability of the typical target and derived a closed-form expression for it using results from the theory of Poisson-Voronoi tessellations. For the more realistic scenario when the obstacle dimensions are finite, our analysis yields an upper bound for the true blind-spot probability of a typical target, provided there exists a sufficiently high intensity of obstacles. Therefore, the asymptotic blind-spot probability can be used to design localization networks such that a typical target's blind-spot probability does not exceed a desired threshold, $\epsilon$.


\bibliographystyle{IEEEtran}
\bibliography{IEEEabrv,JunyangBib}
 
\end{document}